\documentclass[aps,prapplied,twocolumn,superscriptaddress]{revtex4-2}

\usepackage{soul}
\usepackage{amsthm}
\usepackage{amsmath}
\usepackage{amssymb}
\usepackage{graphicx}
\usepackage{epstopdf}
\usepackage{subfig}
\usepackage{multirow}
\usepackage{url}
\usepackage{float}
\usepackage{bm}
\usepackage{ifthen}
\usepackage[usenames,dvipsnames]{color}
\usepackage{mathrsfs}
\usepackage[colorlinks=false,citecolor=blue,urlcolor=black]{hyperref}

\newcommand{\be}{\begin{equation}}
\newcommand{\ee}{\end{equation}}

\usepackage[absolute]{textpos}
\usepackage{xcolor}

\usepackage{placeins}

\newcommand{\sket}[1]{{\ensuremath{\lvert#1\rangle}}}
\newcommand{\lket}[1]{{\ensuremath{\left\lvert#1\right\rangle}}}
\newcommand{\ket}[1]{\if@display\lket{#1}\else\sket{#1}\fi}

\newcommand{\sbra}[1]{{\ensuremath{\langle#1\rvert}}}
\newcommand{\lbra}[1]{{\ensuremath{\left\langle#1\right\rvert}}}
\newcommand{\bra}[1]{\if@display\lbra{#1}\else\sbra{#1}\fi}

\newcommand{\sbraket}[2]{{\ensuremath{\langle#1\rvert#2\rangle}}}
\newcommand{\lbraket}[2]{{\ensuremath{\left\langle#1\!\left\rvert\vphantom{#1}#2\right.\!\right\rangle}}}
\newcommand{\braket}[2]{\if@display\lbraket{#1}{#2}\else\sbraket{#1}{#2}\fi}

\newcommand{\sketbra}[2]{{\ensuremath{\lvert #1\rangle\!\langle #2\rvert}}}
\newcommand{\lketbra}[2]{{\ensuremath{\left\lvert #1\right\rangle\!\!\left\langle #2\right\rvert}}}
\newcommand{\ketbra}[2]{\if@display\lketbra{#1}{#2}\else\sketbra{#1}{#2}\fi}

\theoremstyle{plain}

\theoremstyle{definition}

\begin{document}


\title{Zero-added-loss entanglement multiplexing using time-bin spectral shearing}


\author{Joseph C. Chapman}
\email{chapmanjc@ornl.gov}
\affiliation{Computational Science and Engineering Division, Oak Ridge National Laboratory, Oak Ridge, TN USA}
\author{Muneer Alshowkan}
\affiliation{Computational Science and Engineering Division, Oak Ridge National Laboratory, Oak Ridge, TN USA}
\author{Jack Postlewaite}
\affiliation{Department of Electrical and Computer Engineering, The University of Maryland, College Park, MD, USA}
\author{Saikat Guha}
\affiliation{Department of Electrical and Computer Engineering, The University of Maryland, College Park, MD, USA}
\author{Nageswara Rao}
\affiliation{Computational Science and Engineering Division, Oak Ridge National Laboratory, Oak Ridge, TN USA}



\begin{abstract}
High-quality quantum communications that enable important capabilities, such as distributed quantum computing and sensing, will require quantum repeaters for providing high-quality entanglement. To realize high-rate heralded entanglement for quantum repeaters, Chen \emph{et al.} [Phys. Rev. Appl. \textbf{19}, 054209 (2023)] proposed a scheme for heralded-multiplexed generation of quasi-deterministic entangled photon pairs, called zero-added-loss multiplexing (ZALM). Here, we propose a design of ZALM source using time-bin entanglement and spectral shearing. Additionally, we provide an analysis of experimentally relevant spectral-shearing parameters to optimize the spectral multiplexing. Moreover, we experimentally verify the compatibility of time-bin pulses and spectral shearing, as supported by observation of no {appreciable} phase shift when the same shearing is applied to both time bins. These results expand the benefits of applying a ZALM source to time-bin entanglement use cases. Moreover, more fully demonstrating time-bin and spectral shearing compatibility clears a path towards a broader use of spectral shearing that provides a deterministic frequency shift of high utility.
\end{abstract}


\maketitle

\begin{textblock}{13.3}(1.4,15)\noindent\fontsize{7}{7}\selectfont\textcolor{black!30}{This manuscript has been co-authored by UT-Battelle, LLC, under contract DE-AC05-00OR22725 with the US Department of Energy (DOE). The US government retains and the publisher, by accepting the article for publication, acknowledges that the US government retains a nonexclusive, paid-up, irrevocable, worldwide license to publish or reproduce the published form of this manuscript, or allow others to do so, for US government purposes. DOE will provide public access to these results of federally sponsored research in accordance with the DOE Public Access Plan (http://energy.gov/downloads/doe-public-access-plan).}\end{textblock}

\section{Introduction}
Heralded entanglement distribution is of increasing importance in advancing towards the development of near-deterministic entanglement sources~\cite{PhysRevAppliedZALM2023} and quantum repeaters~\cite{Muralidharan2016}. A spectrally multiplexed heralded entanglement source has been proposed~\cite{PhysRevAppliedZALM2023} which overcomes the limitations of previous sources of multiplexed entanglement that relied on compounded layers of optical switches which grow with the degree of multiplexing~\cite{PhysRevA.84.052326,PhysRevApplied.17.034071}. Chen \emph{et al.}~\cite{PhysRevAppliedZALM2023} propose a zero-added-loss multiplexing (ZALM) where the multiplexing loss is not dependent on the amount of multiplexing being done. To achieve this advantage, frequency multiplexing is used. Spontaneous parametric downconversion (SPDC) is a naturally broadband process, dependent on the crystal phasematching. Ref.~\cite{PhysRevAppliedZALM2023} proposes using broadband SPDC to generate polarization entanglement. In the proposal, two modes, one each from two such entanglement sources, are combined on a beamsplitter with spectrally resolved coincidence detection to enable a spectrally resolved Bell-state measurement. With that measurement information, a feedforward spectral shift is applied so that all heralded photon pairs output with the same frequency. In the case of Ref.~\cite{PhysRevAppliedZALM2023}, and follow-on analysis in Ref.~\cite{PhysRevAppliedZALM2024}, the spectral shift is proposed to be implemented by a non-linear optical up-conversion so the output photons can be collected into a vacancy-center {quantum} memory.

In this work, we take a complementary approach and propose a ZALM source design that leverages time-bin entanglement~\cite{PhysRevLett.62.2205} and uses spectral shearing (a.k.a, serrodyne modulation)~\cite{spectralshearfirst,PhysRevLett.113.053603,Fan2016,PhysRevLett.118.023601} for the spectral shift. Time-bin entanglement is useful for reducing polarization stability requirements of long-links~\cite{Xavier2025} and is more easily compatible with atomic quantum memories which often can only support single-polarization input states. Moreover, spectral shearing provides a deterministic linear frequency shift compatible with single-photons~\cite{PhysRevLett.113.053603,PhysRevLett.118.023601} that does not require additional lasers or optical fields.

Furthermore, in this work, we also provide a detailed optimization analysis for spectral multiplexing based on spectral shearing. In this analysis, we derive the number of multiplexed spectral bins given experimentally relevant spectral-shearing parameters. Using this analysis, we show increased rate of heralded entanglement distribution provided by the multiplexing benefits of a ZALM source using our design. 

A notable concern of our proposal may be that spectral shearing is not obviously compatible with time-bin qubits due to the application of strong phase modulation to the time-bin pulses {which nominally require a steady phase between the pulses}. To validate our proposed ZALM source, we experimentally test the compatibility of time-bin pulses with spectral shearing. First, we provide characterization of our spectral shearing implementation. Then we demonstrate that when the same spectral shearing is applied to both time bins there is no appreciable phase shift between the time bins{, which also signifies no loss of time-bin coherence}. This compatibility was implicit in the results of Ref.~\cite{PhysRevLett.113.053603} but here we provide a more comprehensive verification of such compatibility that can open up additional applications of spectral shearing of time-bin qubits. Related work is done by Ref.~\cite{Nussbaum2023,Nussbaum2026}.

{In Sec.~\ref{sec:ZALMover}, we begin with an overview of ZALM source design. In Sec.~\ref{sec:ZALMoptan}, we present a  multiplexing and generation-rate optimization analysis based on our design choices. In Sec.~\ref{sec:ZALMexpdesign}, we present a detailed description of our proposed experimental ZALM design. In Sec.~\ref{sec:exptbtests}, we describe results from experimental compatibility testing between several aspects of our design, namely time bins and frequency shearing; details on our testing setup are described in App.~\ref{app:tbsetupdesc}. In Sec.~\ref{sec:disc}, we close with a discussion on our results and their implications. }

\section{ZALM design overview}
\label{sec:ZALMover}
\begin{figure}
\centerline{\includegraphics[width=1\columnwidth]{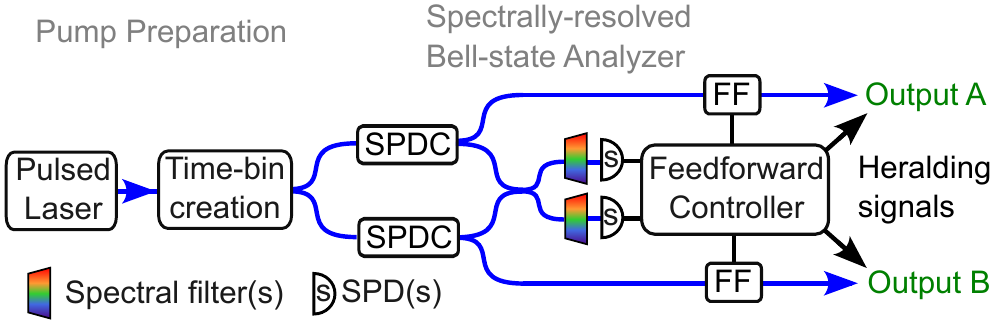}}
\caption{{Simplified experimental setup to produce high-rate heralded entangled photons via the zero-added-loss multiplexing scheme. FF: feedforward. SPD: single-photon detector. SPDC: spontaneous-parametric downconversion crystal.}}
\label{fig:simpZALMsetup}
\end{figure}

{A ZALM source (Fig.~\ref{fig:simpZALMsetup}) relies on spontaneous parametric downconversion (SPDC) emitting photon pairs over a wide bandwidth. Here we show SPDC generation of time-bin entanglement by creating a coherent pair of pump pulses from which the SPDC events could have originated. Within the ZALM source, there are two SPDC sources in an entanglement swapping~\cite{PhysRevLett.80.3891} configuration. This configuration enables heralded entanglement on outputs A-B using detected coincident photons at the output of the Bell-state analyzer beamsplitter. This coincident photon detection provides the desired heralding signal which can be processed and sent to user(s). Using wavelength demultiplexing of some type, the heralding includes information about the photon spectra, often done by discretely binning the continuous output of a frequency or time measurement. For example, the filter(s) transmission profile in the demultiplexing section will define the frequency bins which are being multiplexed together. With this spectral information, the feedforward controller is able to apply the correct feedforward operation on the remaining (now) entangled photons so the output is always in the same spectral mode which works to increase the output rate into that mode via spectral multiplexing.}

{When designing a ZALM source, there are several key choices to be made: (1) What will be used to spectrally resolve the detected photons? (2) What will implement the feedforward spectral shift operation? (3) What type of entanglement is desired?}

{For (1), there are several options including a dispersive time-of-flight spectrometer~\cite{ToF1,ToF2,PhysRevA.105.023708,PhysRevLett.128.063602}, a wavelength-selective switch (WSS)~\cite{WSSJLT2020}, chained fiber Bragg gratings, or integrated optical cavities~\cite{Vahala2003}.}

{Time-of-flight spectrometers operate by leveraging dispersion to provide different arrival times for different wavelengths. This dispersion is a continuous effect but arrival-time binning can be used to enable arrival-time discretization. This spectrometer type has been used for spectrally resolved Bell-state measurement~\cite{PhysRevA.105.023708,PhysRevLett.128.063602}. For a ZALM source, this method has increased loss (about 1 to $>$10~dB) and fiber length (about 1 to $>$10~km) for increased resolution which is undesirable for multiple reasons. Most importantly, to satisfy ZALM's need for feedforward, the output photons would need to be significantly delayed (either by quantum memory or optical delay) to wait for the spectrometer results before feedforward is applied.}

{A wavelength selective switch relies on a dispersive element spatially spreading the bandwidth (i.e., a diffraction grating) and reconfigurable optical system to direct different bandwidth slices to different outputs. In this case, the bins are limited by the output number of the WSS (on the order of 10's of bins) and the spectral resolution of the selectivity (a few GHz). The insertion loss for such a device is fairly fixed but not small at about 5~dB. Overall, this is a pretty solid option for multiplexing on the order of 10's of bins if the loss can not be improved on with other methods.}

{Another option is to chain fiber Bragg gratings with optical circulators since they act as bandpass filters on the reflected light. This is the only method so far that can reasonably provide frequency-bin widths $<$1~GHz. It also has the benefit of being simple and robust with just several discrete fiber-based components. But is it not actually zero-added loss for increased bins since the chain would get longer.} This can be viewed as somewhat of a design trade-off compared to the cascaded sources in Ref.~\cite{PhysRevApplied.17.034071}. In that context, this method essentially leverages multiple sources, each at a different spectral mode of the same broadband SPDC source instead of a single-spectral mode from multiple SPDC sources. In this way, there is not truly no added loss due to extra multiplexing but it is not on the output photon paths (so only reduces the heralding rate). The filter order could be optimized to give preferential treatment to certain bins (near the center) if desired.

{To avoid the loss of chaining discrete filters via circulators, integrated optical cavities could be used which are connected via a bus waveguide~\cite{766787, Yang14PIC}. Here the loss due to chaining is primarily from the waveguide propagation loss ($<$1~dB/cm), which can support many micro-resonators in a short distance. This has the added benefit that very narrow bandwidths could be filtered using high-Q cavities but care is required not to transmit more than 1 cavity resonance. Additionally, high-efficiency chip-fiber coupling is required or the benefit of the on-chip components could be offset by the losses to get on and off chip. }

{In consideration of a low-loss robust relatively easy first demonstration of our proposed ZALM source, in our design for (1) we chose the chained fiber Bragg gratings. If additional loss (and higher cost) is tolerable, then a wavelength-selective switch is a fairly easy drop-in replacement.}

{For (2), a quantum-state preserving frequency shift is required. Frequency conversion, electro-optic modulation, and acousto-optic modulation are several primary methods to implement that shift. In the other ZALM literature~\cite{PhysRevAppliedZALM2023,shapiroZALM2025}, mode conversion~\cite{PhysRevA.101.042322, PhysRevLett.124.160501} and quantum frequency conversion (QFC) are chosen respectively because there the focus is to couple into a vacancy-based quantum memory where QFC from telecom photons is already required. In this case, the pump or the input photons need to be altered to achieve a spectral shift of the input photons. Directly shifting laser wavelengths is usually accomplished on time scales $>1~\mu s$ (via piezo or thermal tunings). Frequency comb lines or multiple pump lasers could be switched to use as pump but this would be undesirable to need multiple lasers. The aforementioned, mode conversion can be used for optically controlled mode conversion to different frequency modes of a resonator.}

{Alternatively, Acousto-optic modulators can be used to provide varying shifts but are often limited to bandwidths on the order of 10-100~MHz due to the frequency dependence of the diffraction angle. Moreover, electro-optic phase modulators with a serrodyne (i.e., frequency shearing) drive signal~\cite{spectralshearfirst,PhysRevLett.113.053603,Fan2016,PhysRevLett.118.023601} consisting of a voltage ramp, have been shown to produce very linear frequency shifts on the order of THz, compatible with single-photons~\cite{PhysRevLett.113.053603,PhysRevLett.118.023601}. Notably, these frequency shifting techniques can be stacked to enable varying performance tailored to the system needs, i.e., applying frequency shearing then quantum frequency conversion or quantum frequency conversion then mode conversion~\cite{PhysRevAppliedZALM2023}.}

{Finally for (3), one can consider different types of entanglement as varied as the degrees of freedom of the photon carrying it (e.g., polarization, spatial mode, angular momentum, path, frequency, time-bin, and temporal mode). A common choice in this networking context is polarization. This is considered in other ZALM literature~\cite{PhysRevAppliedZALM2023,shapiroZALM2025} and has the benefits of being largely insensitive to Doppler shifts of the satellite-ground link considered in Ref.~\cite{PhysRevAppliedZALM2023}. Although polarization encoded entanglement requires dynamic polarization compensation needed to stabilize the polarization channel experienced by the entangled photons. Note this polarization channel stabilization is a more stringent criterion than just stabilizing the state of polarization~\cite{Chapman24APC}. In the satellite context, these polarization rotations is primarily from geometrical rotations of the satellite and ground station but in fiber optics, it is more dependent on various environmental factors. Alternatively, another common encoding is time bins where the basis states are optical pulses separated by a fixed time-bin separation. This state is less sensitive to polarization fluctuations but sensitive to a Doppler shift on the time-bin spacing~\cite{PhysRevApplied.14.014044}. This encoding also has the advantage of being more directly compatible with atomic system which often have optical transitions that are polarization sensitive. In our analysis, we focus on time-bin entanglement due to these advantages and it is unexplored in the ZALM context.}

\section{ZALM spectral multiplexing of time bins optimization Analysis}
\label{sec:ZALMoptan}
Here we provide a model for spectral multiplexing of time-bin qubits allowing optimization of experimentally relevant parameters for maximum multiplexing performance. Moreover, we also analyze the noise resilience of frequency shearing in applying unwanted phase shifts to the signal in addition to the frequency shift. Finally, using our multiplexing model, we examine useful parameter regimes as well as comparing heralded entanglement rates for different multiplexing configurations.
\label{sec:optanalysis}
\subsection{Model of spectral multiplexing of time bins}
To model the multiplexing of the ZALM source, we first define the frequency-bin width (using full width at half-maximum, FWHM) $\delta f_b$. Any frequency-bin filter will also have a finite transition band from the pass-band to the stop-band which is represented as an additional spacing $\delta_s$. From the spectral profile of the frequency-bin filter, using the Fourier transform, the time-bandwidth product (TBP) ${P_{T,B}=}\Delta f\Delta t$ of that spectral mode can be calculated. Using the above, to maximize frequency bin distinguishability, we propose a frequency-bin spacing of 
\begin{equation}
    \Delta F_b = \frac{\delta f_b}{{P_{T,B}}} + \delta_s. \label{eq:DelFb}
\end{equation}
Dividing by the TBP adds a variable amount of additional spacing based on the spectral profile. Comparing the TBP of a flat top ${P_{T,B}}=0.89$ with that of a Gaussian ${P_{T,B}}=0.44$ and Lorentzian ${P_{T,B}}=0.17$ shows that as the wings of the spectral distribution spread out more the TBP goes lower which is why we include the TBP in the denominator. Using the TBP, we also specify the Fourier-conjugate pulse width for the frequency bin
\begin{equation}
    \tau_b = \frac{{P_{T,B}}}{\delta f_b}.
\end{equation}
For time-bin qubits, it is important to have sufficient separation so there is not interference between the modes when passing through a single UMZI. By analyzing the calculated interference of two shifted Gaussians, a shift of $> 10$ FWHM (or about $>$24 standard deviations $\sigma$) is needed to ensure $<<$1\% interference. As a minor approximation, we use this rule of thumb for all TBP modeled resulting in a time-bin spacing
\begin{equation}
\Delta t_b = \frac{24}{2.355} \tau_b.
\end{equation}
Moreover, to ensure high-fidelity frequency shifting via spectral shearing, the entire pulse should fit within the time of the spectral shearing ramp waveform. To ensure it is well contained within that ramp and the tails of the pulse are not seeing the peaks of the ramp waveform, the spectral-shearing drive frequency is constrained by the frequency-bin pulse width. This varies for different waveforms because some of the same slope sign for the entire waveform (sawtooth) whereas others change sign for half of the waveform (sine and triangle) which leads to needing to constrain the pulse to fit in half the waveform. Thus, for these waveforms, the spectral shearing drive frequency is upper bounded by
\begin{align}
    D_A &\leq \frac{1}{\frac{8}{2.355} \tau_b}=\frac{3}{\Delta t_b} \text{ (Sawtooth),}\\
    D_T &\leq \frac{1}{2\frac{8}{2.355} \tau_b}=\frac{3}{2\Delta t_b} \label{eq:DTub}\text{ (Triangle), and}\\
    D_S &\leq \frac{1}{2\frac{8}{2.355} \tau_b} =\frac{3}{2\Delta t_b}\label{eq:DSub}\text{ (Sine)}.
\end{align}
Additionally, $D$ should be an integer multiple of $1/\Delta t_b$ so that both time bins always match up with the same part of the spectral shearing waveform. The maximum spectral shearing drive frequency meeting these constraints is 
\begin{align}
    D_A &=\frac{3}{\Delta t_b} \text{ (Sawtooth),}\\
    D_T &=\frac{1}{\Delta t_b} \text{ (Triangle), and}\\
    D_S &=\frac{1}{\Delta t_b}\text{ (Sine)}.
\end{align} 
Moreover, a pump repetition rate can be calculated such that the time bins align with the spectral shearing waveform but time bins from successive pump pulses do not overlap after going through the analyzer interferometer. The pump repetition rate is $R_P=1/(3\Delta t_b)=D_A/9=D_T/3=D_S/3$.

The magnitude of the frequency shift from spectral shearing is dependent on the slope. For a given waveform, assuming an average-power-limited amplifier is used, the peak voltage  can be calculated from the power using root-mean-square (RMS) integration over the waveform shape assuming a 50-Ohm impedance. For an average power $P$ (Watts), the peak voltage is
\begin{align}
    V_A &= \frac{\sqrt{50 P}}{0.585382} \text{ (Sawtooth),}\\
    V_T &= \frac{\sqrt{50 P}}{0.579814} \text{ (Triangle), and}\\
    V_S &= \frac{\sqrt{50 P}}{\frac{1}{\sqrt{2}}} \text{ (Sine)}\label{eq:Vs}.
\end{align}
According to Ref.~\cite{PhysRevLett.119.083601}, for a given slope $\mathcal{A}$, the frequency shift from spectral shearing will be
$\Delta f_s = \frac{\mathcal{A}}{2 V_{\pi}},$

where $V_{\pi}$ is the $\pi$-voltage of the phase modulator. Thus, for the waveforms in question, the expected frequency shift is 
\begin{align}
    \Delta f_A &= \frac{V_A D_A}{V_{\pi}} \text{ (Sawtooth),}\label{eq:DelfA}\\
    \Delta f_T &= \frac{V_T 2 D_T}{V_{\pi}} \text{ (Triangle), and}\label{eq:DelfT}\\
    \Delta f_S &= \frac{V_S \pi \cos{(\phi_S)} D_S}{V_{\pi}} \text{ (Sine)}\label{eq:DelfS},
\end{align}
where $\phi$ is the phase of the sine wave with respect to the optical pulse being sheared. Using this frequency shift and the bin spacing, the number of bins that can be multiplexed together is
\begin{align}
    n_A &= 2\frac{\Delta f_A}{\Delta F_b} + 1= 2\frac{28.44697 \delta f_b \sqrt{P} }{\sigma V_{\pi} {P_{T,B}} (\frac{\delta f_b}{{P_{T,B}}} + \delta_s) } + 1 ,\label{eq:nA}\\
    n_T &= 2\frac{\Delta f_T}{\Delta F_b} + 1= 2\frac{19.14679 \delta f_b \sqrt{P} }{\sigma V_{\pi} {P_{T,B}} (\frac{\delta f_b}{{P_{T,B}}} + \delta_s) } + 1,\label{eq:nT}\\
    n_S &= 2\frac{\Delta f_S}{\Delta F_b} + 1= 2\frac{24.66150 \delta f_b \sqrt{P} }{\sigma V_{\pi} {P_{T,B}} (\frac{\delta f_b}{{P_{T,B}}} + \delta_s) } + 1\label{eq:nS},
\end{align}
for the sawtooth, triangle, and sine waves, respectively.
\begin{figure}
\centerline{\includegraphics[width=1\columnwidth]{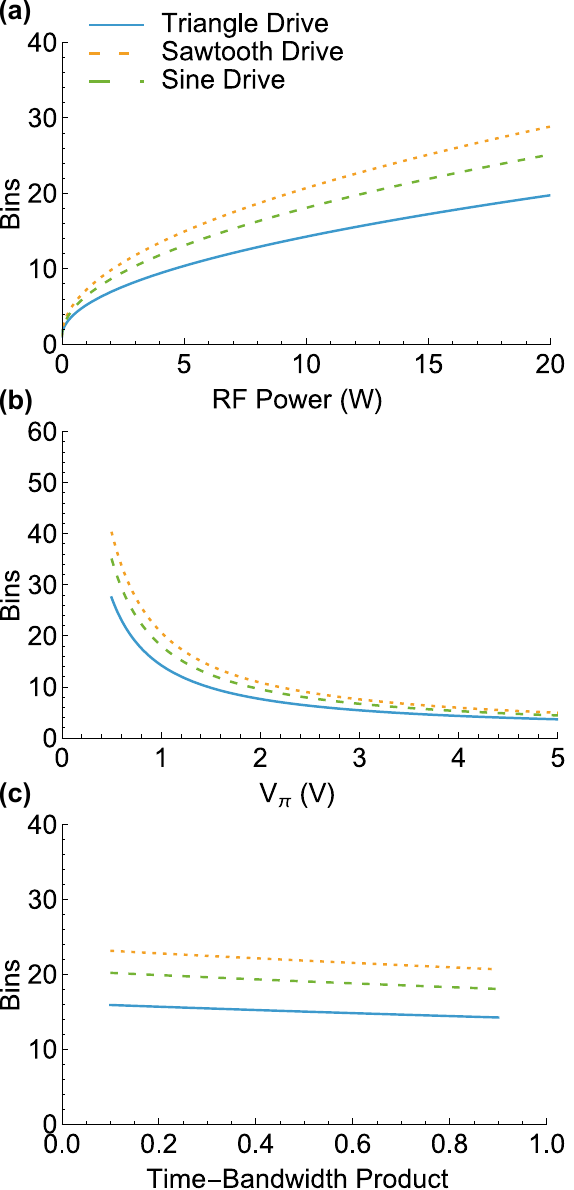}}
\caption{Spectral-shearing parameter analysis for frequency-bin multiplexing optimization. Maximum number of allowed frequency bins versus spectral shearing (a) input RF power, (b) modulator $V_{\pi}$, and (c) input pulse time-bandwidth product. The different curves on each plot compare spectral-shearing drive waveform shapes: triangle wave, sawtooth wave, and sine wave. If not the variable along the X-axis, the simulation parameters used are: $P=10$~W; $V_{\pi}=1$~V; $\delta f_b=12.5$~GHz; $\Delta F_b=2$~GHz; ${P_{T,B}} = 0.89$.}
\label{fig:nbinsvspowpitbp}
\end{figure}

 \subsection{Spectral-shearing voltage-noise analysis}
Here we analyze the {adverse e}ffect of voltage noise on spectral shearing. To start, the electric field entering the phase modulator (frequency shifter) is:

\begin{equation}
    E_{in}(x,t)=|E_0(x,t)|e^{i(\omega_{s}t-kx)}
\end{equation}

With carrier $\omega_s$, we will apply a frequency shift with a linear scaling voltage over the entire pulse duration. Applied signal $V(t)=\mathcal{A}t+\delta V$ which includes small (constant) voltage offset. When applied to phase modulator with characteristic $V_\pi$, the overall applied phase is then:

\begin{equation}
    \phi(t)=\frac{\pi\mathcal{A}t}{V_\pi}+\frac{\pi\delta V}{V_\pi}
\end{equation}

And we can write our field, after the $\tau$ duration of applied phase signal, as:

\begin{equation}
    e^{i\phi(t)}E_{in}(x,t)=|E_0(x,t)|e^{i([\omega_s+\frac{\pi A}{V_\pi}]t+\frac{\pi \delta V}{V_\pi}-kx)}
\end{equation}

So for a frequency shift of $\Delta\omega=\frac{\mathcal{\pi A}}{v_\pi}$, we also pick up a phase change of $\Delta\phi=\frac{\pi\delta V}{V_\pi}$. For a DC offset voltage, the maximum allowable voltage noise ($V_{RMS}=V_{Peak}$) for an allowable phase shift is then:

\begin{equation}
    \delta V = \frac{\Delta\phi  V_\pi}{\pi}
\end{equation}

For  $V_\pi=1$~V and $\Delta\phi=5^{\circ}$ then $\delta V= 28$~mV. This provides a quantification of allowable voltage noise on the shearing waveform. Current technology can produce waveforms with less noise providing a high-quality waveform for frequency shearing. A better modeling of this noise should consider that the noise signal is carried on harmonics of the base signal and will have functional dependence on the currently applied linear scaling signal.

\subsection{Spectral multiplexing modeling results and analysis}
 Comparing Eq.~\eqref{eq:nA}, Eq.~\eqref{eq:nT}, and Eq.~\eqref{eq:nS}, we see the differences between the waveforms distinguish themselves for how much frequency multiplexing they can enable. This is primarily because the expected frequency shifts are all different because of waveform shapes as seen in Eq.~\eqref{eq:DelfA}, Eq.~\eqref{eq:DelfT}, and Eq.~\eqref{eq:DelfS}. In addition, $D_A>D_T=D_S$ which further contributes to different realized frequency shifts. Even with its higher RMS factor in the denominator of Eq.~\ref{eq:Vs} (which produces a smaller frequency shift), the local slope of the sine wave at the zero crossing is significantly higher than a comparable triangle wave, evidenced by the factor of $\pi$ in Eq.~\eqref{eq:DelfS} compared to the factor of 2 in Eq.~\eqref{eq:DelfT} so that for maximum spectral shearing the sine wave is the better waveform choice of the two but still lags behind the sawtooth wave because $D_A=3D_S$ and $V_A>V_S$.
 
In Fig.~\ref{fig:nbinsvspowpitbp}, we compare the calculated number of bins for the candidate waveforms as a function of the RF power, modulator $V_{\pi}$, and TBP. It is clear that high power and low $V_{\pi}$ are crucial for multiplexing many bins. The lower time-bandwidth product makes a modest improvement which accords with  Eq.~\ref{eq:nA}, Eq.~\ref{eq:nT}, and Eq.~\ref{eq:nS} where the TBP appears twice and partially cancels itself out in the denominator.

\begin{figure}
\centerline{\includegraphics[width=1\columnwidth]{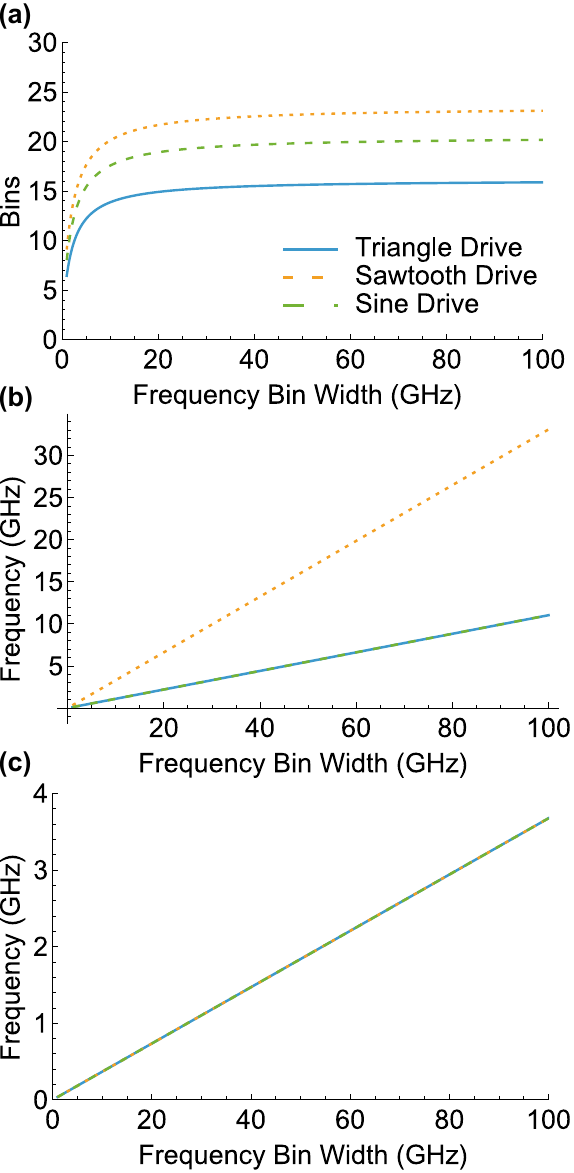}}
\caption{Frequency-bin width analysis for spectral-shearing multiplexing. (a) maximum number of allowed frequency bins $n$ (b) spectral shearing drive frequency $D$ (c) pump repetition rate $R_p$ versus frequency-bin width $\delta f_b$. The different curves on each plot compare spectral-shearing drive waveform shapes: triangle wave, sawtooth wave, and sine wave. The simulation parameters used are: $\Delta F_b=2$~GHz; ${P_{T,B}} = 0.89$.}
\label{fig:nbinsfreqshrRRvsFBWid}
\end{figure}

Similarly, Fig.~\ref{fig:nbinsfreqshrRRvsFBWid}(a) shows only a slight increase in number of multiplexed bins for a wider frequency-bin width after 20 GHz when the added spacing $\delta_s=2$~GHz becomes relatively negligible. Fig.~\ref{fig:nbinsfreqshrRRvsFBWid}(b)-(c) show that optimal spectral shearing drive frequency and the pump repetition rate, respectively, vary linearly with frequency bin width. Moreover, the sine wave and triangle wave have the same optimal frequencies whereas the sawtooth wave can have three times the drive frequency but they all have the same repetition rate. A higher repetition rate is good for increasing the rate of entanglement distribution because it scales linearly with the rate of entanglement as we will discuss more below. It is important to note that a sinewave being a single frequency component can operate at higher frequencies with lower bandwidth equipment compared to the other waveforms. The sawtooth and triangle waves have many higher frequency components, so for a given frequency, they need more than ten times the analog bandwidth compared to the sine wave to produce a high-quality waveform.
\begin{figure}
\centerline{\includegraphics[width=1\columnwidth]{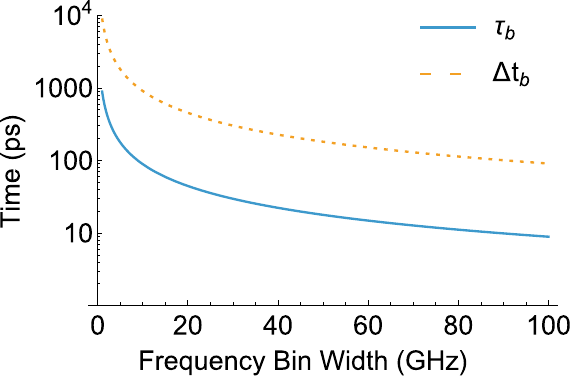}}
\caption{Spectral shearing time-domain analysis versus frequency-bin width. The simulation parameters used are: $\Delta F_b=2$~GHz; ${P_{T,B}} = 0.89$.}
\label{fig:timebinvsFBWid}
\end{figure}

We now turn to the time domain. In Fig.~\ref{fig:timebinvsFBWid}, we indeed see the optimal time-bin pulse width and time-bin spacing separated by a factor of 10. Moreover, we see the pulse width and spacing decrease as the frequency bin size increases due to the time-bandwidth product of a Fourier conjugate pulse. This means that the frequency-bin choice is tied to very practical considerations for pump-laser choice and environmental stability. Laser technology varies as to what optical pulse widths can be produced. Mode-locked lasers can produce the shortest pulses, easily reaching pico-second and femto-second timescales. Whereas, gain-switched lasers and amplitude-modulated CW lasers are generally limited to a minimum around 30-50 ps. Amplitude-modulated CW lasers have the most flexibility in their pulse width, only dependent on the electronic drive, assuming the modulator has enough bandwidth. Mode-locked lasers benefit from the highest optical extinction of light in between pulses which is important for time-bin qubits to not experience unwanted interference. Whereas, both gain-switched and amplitude-modulated CW lasers need careful bias tuning to ensure good inter-pulse extinction and may require additional stages of modulation to increase extinction.

Moreover, the pulse width also determines the necessary stability of the path lengths going to the Bell-state measurement and spectral shearing. Fiber can drift in path length considerably over long distances~\cite{PhysRevApplied.21.014024} but even over short ones if the temperature is not stable meaning that if the drift is comparable to, or more than, the pulse width, path-length stabilization will be required leading to additional complication. In our design, we elected for 12.5-GHz bins in part due to the fact that the frequency-bin pulse widths would then be big enough to avoid needing active path-length stabilization in our lab.

\begin{figure}
\centerline{\includegraphics[width=1\columnwidth]{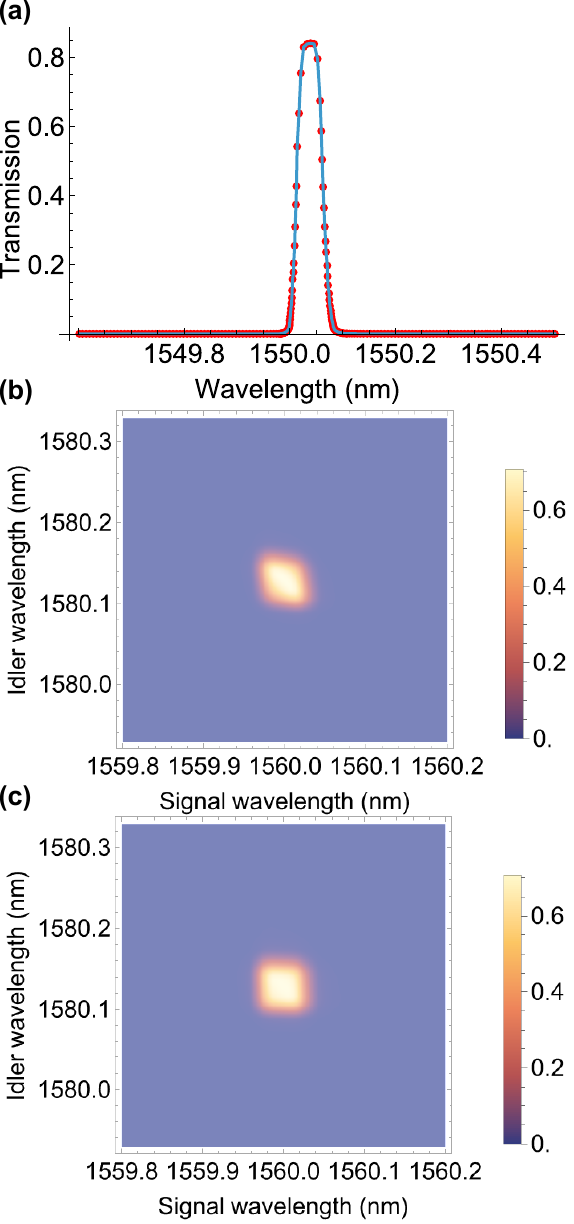}}
\caption{Joint-spectral analysis with filtering and pump bandwidth variation {using $200 \times 200$ sampling grid}. Color bar units arbitrary. (a) Representative fiber-Bragg grating filter function for 12.5-GHz channels (O/E Land) and joint-spectral amplitude filtered as with the filter shown in (a) on each marginal for SPDC source modeled with Gaussian pump-envelop{e} function using (b) 70-ps pump pulse duration and 12.9-GHz bandwidth and (c) 35-ps pump pulse duration and 25.8-GHz bandwidth and Sinc phase-matching function for periodically poled type-0 lithium niobate.}
\label{fig:JSA}
\end{figure}

The optimal frequency-bin pulse width does not directly determine the pump pulse width as is seen from Fig.~\ref{fig:JSA}. The frequency-bin filter is applied to the marginal distribution along the X- or Y-axis. Whereas, the pump pulse width contributes to the pump-envelope function width which goes across the diagonal due to energy conservation. Thus, to ensure no appreciable frequency entanglement which contributes to distinguishability and reduces entanglement swapping fidelity~\cite{PhysRevA.64.063815}, Fig.~\ref{fig:JSA} shows that it is important to have a pump pulse width shorter than the optimal frequency-bin pulse width so that the pump spectrum (along the diagonal) overfills the frequency-bin filter (on the marginal). Through our calculations, we found that if the frequency-bin pulse width (FWHM) and the pump pulse width (FWHM) are the same then the purity is about 0.95, whereas when the pump pulse width is half the frequency-bin pulse width, the purity is about 0.995.

\begin{figure}
\centerline{\includegraphics[width=1\columnwidth]{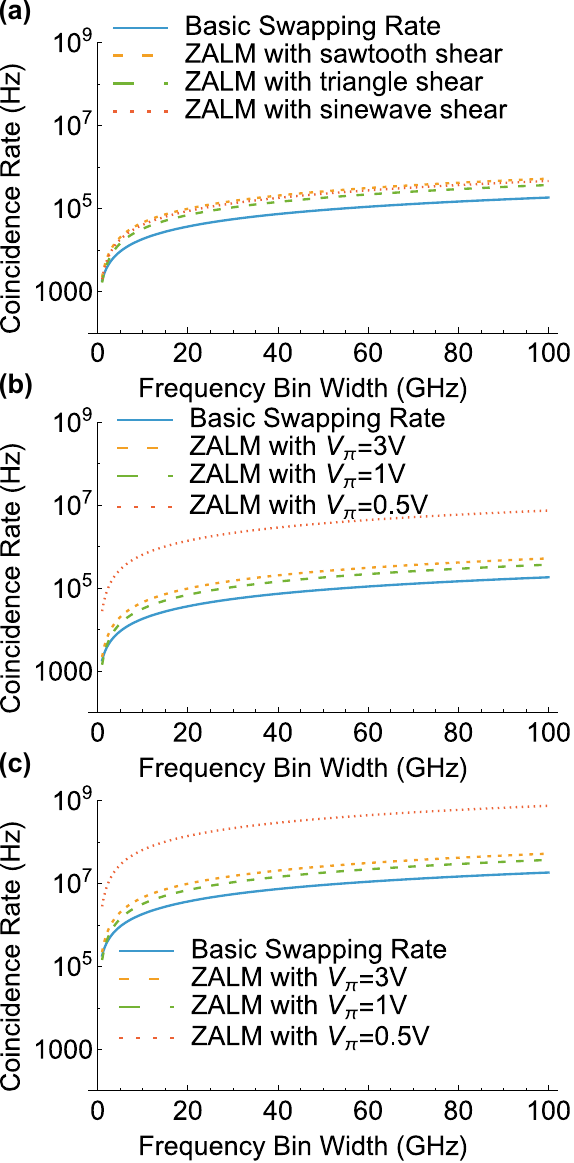}}
\caption{Heralded coincidence rates comparing a basic heralded entanglement source (from entanglement swapping) to a ZALM source with different system parameters: (a) waveform shape with pair-production probability $P_p=0.01$. (b) spectral-shearing phase modulator with $P_p=0.01$ for $V_{\pi}=3$~V with insertion loss $L_I=3$~dB, $V_{\pi}=1$~V with $L_I=6$~dB, and $V_{\pi}=0.5$~V with $L_I=1$~dB. (c) Same as (b) except $P_p=0.1$.}
\label{fig:CoincRates}
\end{figure}

Finally, putting it all together, in Fig.~\ref{fig:CoincRates} we examine estimated entanglement rates for the ZALM source under different design parameters{. We start by} assuming a simplified pulsed SPDC model, i.e., $C_{AB} = R_p P_p \eta_A \eta_B$ {where the coincident rate $C_{AB}$ is equal to the laser pulse repetition rate $R_p$ multiplied by a pair production probability $P_p$ and the channel loss of each mode $\eta_i$~\cite{chapphowest2018}. Since SPDC is known to more generally produce states which sum over multiple pairs~\cite{takesue2010effects}, this is a first order model assuming $P_p<<1$~\cite{chapphowest2018}. Moreover, we assume a} 50\% Bell-state measurement efficiency{~\cite{PhysRevA.51.R1727}}.

Fig.~\ref{fig:CoincRates}(a) illustrates the impacts of the different spectral shearing waveforms on the overall heralded coincidence rate. The rate using the sine wave shearing is so close to the sawtooth wave that given the additional bandwidth (and thus significant increased equipment costs) required for the sawtooth waveform, in most cases the sinewave shape is the better choice unless the application requires the best achievable performance with little care for cost. This illustrates why it is important to take a system engineering approach and model the system from beginning to end otherwise various competing factors may be mis-categorized in their overall contributions.

Next,  in Fig.~\ref{fig:CoincRates}(b)-(c), different phase modulator options are compared due to strong dependence of bins $n$ on $V_{\pi}$ shown in Fig.~\ref{fig:nbinsvspowpitbp}(b). Here the $V_{\pi}=3$~V devices is attributed insertion loss $L_I=3$~dB and the $V_{\pi}=1$~V devices is attributed $L_I=6$~dB to compare between conventional and low-$V_{\pi}$ thin-film lithium-niobate modulators available. These specifications are approximately what is found by vendors in this market. The thin-film device is attributed to lower fiber-coupling e{fficiency and more propagation loss} due to its {mode size and} longer length which is how it achieves the low $V_{pi}$. We also add a hero device, outside of current technological capabilities, with $V_{\pi}=0.5$~V and $L_I=1$~dB as an example of how beneficial modulator improvements would be to this scheme.

For the modulators on the market, Fig.~\ref{fig:CoincRates}(b) shows it is better to have lower insertion loss than lower $V_{\pi}$ due to the $V_{\pi}=3$~V having the higher rate over the $V_{\pi}=1$~V device. But that is a minor difference compared to the improved rate of the hero device which clearly shows that lower $V_{\pi}$ and lower insertion loss together make a big difference. Moreover, Fig.~\ref{fig:CoincRates}(c) shows the highest rates owing to $P_p=0.1$ compared to $P_p=0.01$ in Fig.~\ref{fig:CoincRates}(b). At $P_p=0.1$ and higher it would definitely be beneficial to use photon-number-resolving (PNR) detection at the Bell-state measurement due to the considerable increase in multiple-pair events. As shown in Ref.~\cite{PhysRevApplied.17.034071}, when PNR detection is used, the ZALM source benefits from additional resilience to multiple-pair events due to its design which can lead to higher rates with better entanglement quality.

\section{ZALM experimental design description}
\label{sec:ZALMexpdesign}
\begin{figure*}
\centerline{\includegraphics[width=1\textwidth]{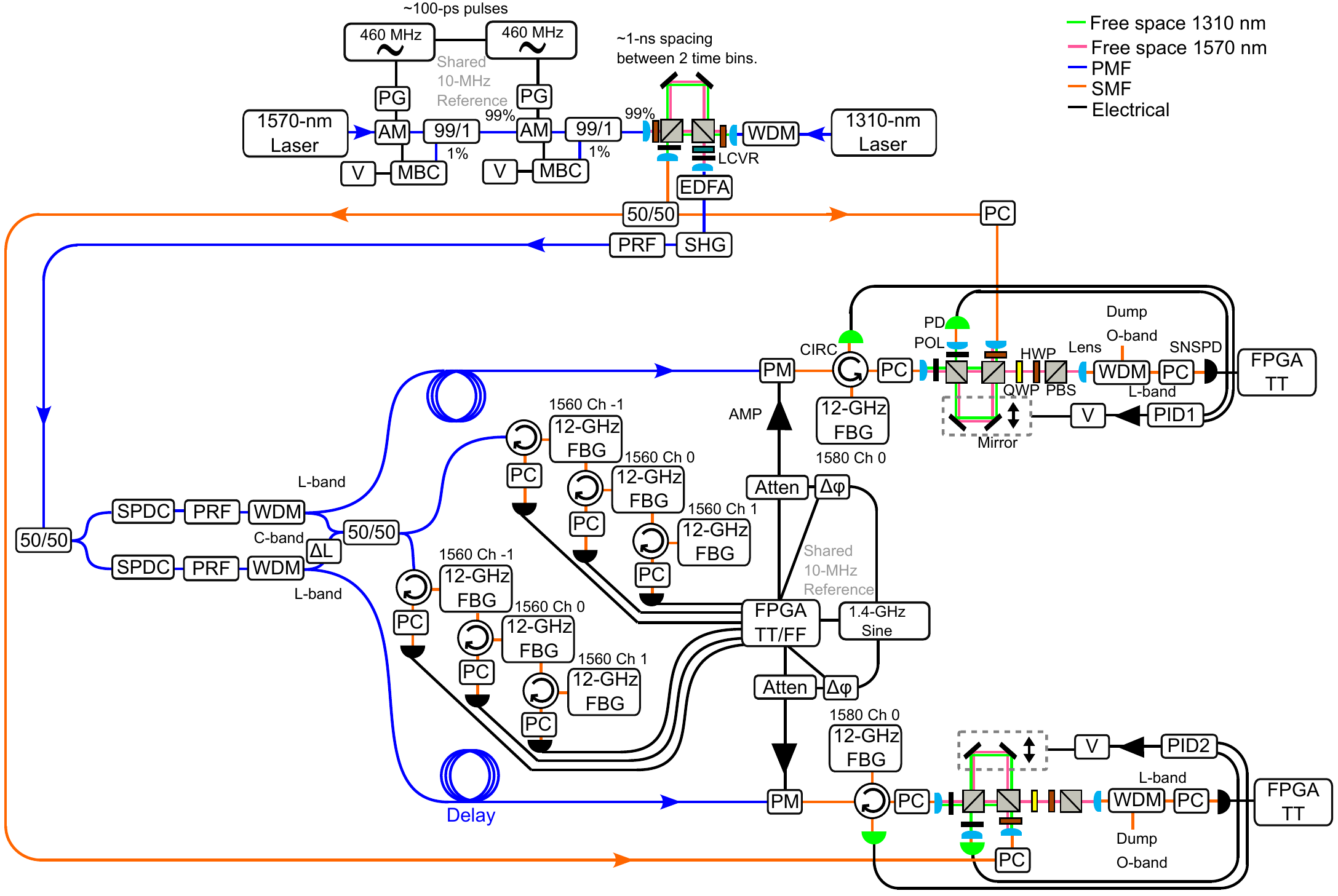}}
\caption{Proposed experimental setup to produce high-rate heralded entangled photons via the zero-added-loss multiplexing scheme. AM: amplitude modulator. AMP: RF amplifier. Atten: Attenuator. CIRC: fiber-optic circulator. EDFA: erbium-doped fiber amplifier. FF: feedforward. FBG: fiber-Bragg grating. FPGA: field-programmable gate array. HWP: half-wave plate. LCVR: liquid-crystal variable retarder. MBC: modulator-bias controller. PBS: polarizing-beamsplitter. PC: polarization controller. PD: photodiode. PG: pulse generator. PID: proportional-integral-derivative controller. PM: phase modulator. PMF: polarization-maintaining fiber. POL: polarizer. PRF: pump-reject filter. QWP: quarter-wave plate. RF: radio frequency. SHG: second-harmonic generation crystal. SMF: single-mode fiber. SNSPD: superconducting-nanowire single-photon detector. SPDC: spontaneous-parametric downconversion crystal. WDM: wavelength-division multiplexing. V: DC voltage source.}
\label{fig:ZALMsetup}
\end{figure*}

Here we present the results of our experimental design to implement a ZALM source for time-bin qubits using spectral-shearing-based multiplexing as summarized in Fig.~\ref{fig:ZALMsetup}. 

A fiber-coupled continuous-wave laser is sent through two stages of DC-bias controlled amplitude modulation. The amplitude modulators are driven by synchronized pulse generators to produce a high-extinction ratio optical pulse train of approx. 100 ps with a repetition rate of about 460 MHz. The pulsed laser is then directed into a polarization-adjusted unbalanced Mach-Zehnder interferometer (UMZI) to create the pump time bins. Another laser, at a different wavelength, also goes through the interferometer to serve as the phase stabilization reference {to stabilize the relative phase between the time bins through the optical system (including another delay interferometer for analysis/measurement)}. The polarization adjustment serves to enable easy balancing of the optical power in each time bin; a liquid-crystal is added after the interferometer for easy pre-adjustment of the time-bin phase. The two-stages of amplitude modulation are useful to ensure there is not interference from the pump interferometer; alternatively, a single-stage of amplitude modulation of sufficiently high extinction could be used such that there is not appreciable interference of light between the short and long paths of the pump interferometer. The output of the pump time-bin interferometer is amplified and directed toward a second-harmonic-generating crystal after which the (longer-wavelength) pump is filtered out. The remaining light is used to pump two spontaneous parametric downconversion (SPDC) sources. Here we choose to have the center of the SPDC spectrum straddle the optical C- and L-band so that one of the photons in the pairs is in each band. Due to the flexibility of periodic poling and wide availability of lasers and filters, one could easily choose other wavelengths. The SPDC photons are filtered so the C-band photons go in one fiber and the L-band photons go in another. 

The C-band photons meet on a 50/50 beamsplitter after adjusting for any residual delay mismatch from the SPDC source. The output modes of the beamsplitter are then sorted into frequency bins using fiber-Bragg gratings (FBG) filters and optical circulators. {As an example, in Fig.~\ref{fig:ZALMsetup}, we show three bins; based on the analysis in Fig.~\ref{fig:nbinsvspowpitbp}, this frequency-shearing design can support 13 bins and 18 bins at 5~W and 10~W RF power, respectively.  The filter} outputs are then detected with single-photon detectors; superconducting nanowire single-photon detectors are recommended due to their low-noise, low-timing-jitter, high-efficiency capabilities. The outputs of which are directed towards a time-tagging feedforward controller (likely a field-programmable gate array device) which processes coincident detection events within frequency bins, saving their time stamps, while also producing dynamic output control signals to the frequency shift stage for every detection event. This requirement means the feedforward controller must be fast enough to handle this processing for every pump pulse likely requiring high-clock-rate pipelined digital logic and high-speed frequency shift electronic control.

In the other mode, the L-band photons are transmitted through a delay and then into a electro-optic phase modulator. This phase modulator is driven with a approx. 1.4-GHz sine wave (rationale explained in future sections) with the phase calibrated to be at the zero crossing when the time bins are in the modulator. To affect a certain frequency shift on the time bins, the feedforward controller dynamically adjusts the slope of the sinewave using a fast reprogrammable radio-frequency (RF) attenuator. Additionally the slope sign can be adjusted by applying a 180$^{\circ}$ phase shift to the sinewave either to the generator itself or by using an external delay-based fast phase shifter (could be constructed simply with two fast RF switches and appropriate-length delay cables). The frequency shift is chosen to shift all other frequency bins to the zero-th bin in the center of the SPDC spectrum. After the phase modulator, a frequency-bin filter is used to only transmit photons in the zero-th bin.

The resultant entangled photon pairs can then be used for their desired application. At this point, a source of spectrally pure heralded entangled photons is produced. To determine the state quality, we show the photons being directed to projective measurement setups for tomographic state analysis. First, a time-bin-to-polarization qubit converter~\cite{PhysRevA.66.040301,victora2020new} is used followed by a conventional wave-plate-based polarization measurement subsystem. The time-bin phase-reference light is used to stabilize the phase of the unbalanced interferometer as part of the time-bin-to-polarization converter. Following the polarization measurement, there is a filter to remove any contaminating stabilization photons. Following which, the entangled photons are detected using single-photon detectors whose outputs are time-tagged for coincidence analysis.

\section{Experimental testing of spectral shearing compatibility with time-bins}
\label{sec:exptbtests}
To this point, we have presented design and modeling for a proposed ZALM source implementation. Our design hinges on the compatibility of spectral shearing with time-bin qubits. Due to the {Fourier relationship} between frequency noise and phase noise in combination with the large frequency shifts applied during spectral shearing {using a waveform driving a \emph{phase} modulator}, it was not at all obvious to the authors at the outset that spectral shearing would be compatible with time-bin qubits {that require a steady \emph{phase} relationship between the pulses. As a result, if the frequency shearing applied is not identical for both time bins, that would cause a relative phase shift or phase noise.} 
Thus, here we describe our results of experimentally testing that compatibility to build confidence in this aspect of the design. {Notably, this confidence can extend beyond preserving the time-bin phase because if the phase is preserved, so is the coherence. This is because a loss of coherence is often caused by excessively large phase shifts that result in dephasing due to finite coherence lengths.} 

Experimentally, for this testing we sent a pulsed laser through a delay-line interferometer then applied some varied type of spectral shearing to both pulses and sent that signal through  {another delay-line interferometer} for phase-shift analysis{, i.e., the analyzer interferometer}. We also examined the optical and electrical signals coming into the spectral-shearing modulator. We describe our test setup in more detail in Appendix~\ref{app:tbsetupdesc}. For these tests, we used classical power levels but as has been shown before for spectral shearing, these results will directly translate to single-photons~\cite{PhysRevLett.113.053603,Fan2016,PhysRevLett.118.023601}.

\begin{figure}
\centerline{\includegraphics[width=1\columnwidth]{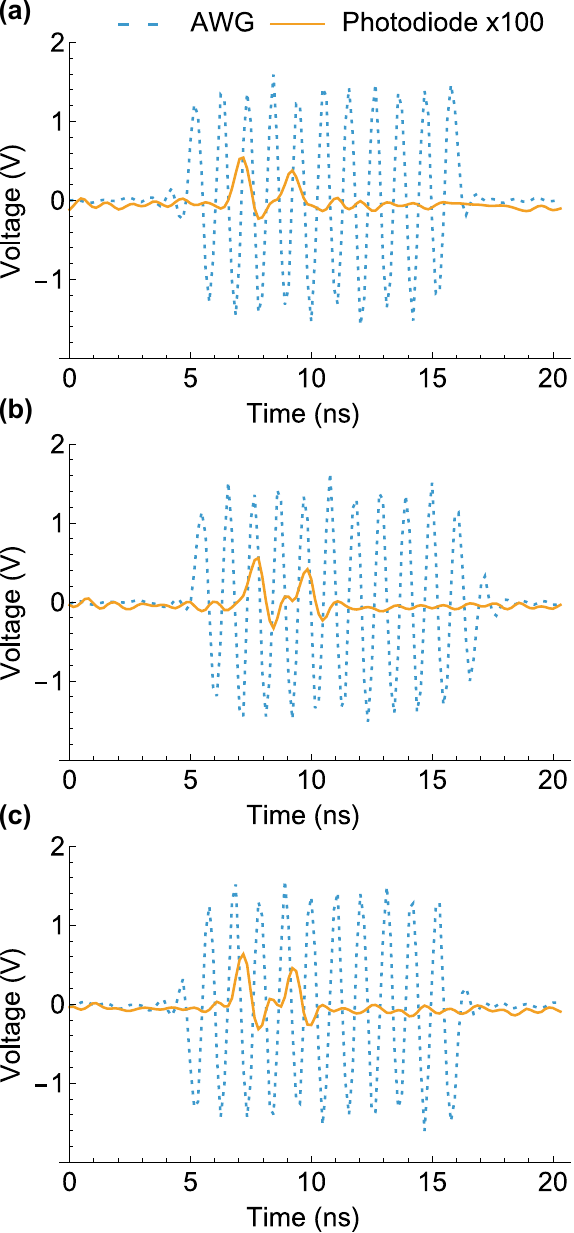}}
\caption{Measured classical time-bin signals entering spectral shearing phase modulator for the sine wave shearing waveform. The phase between the optical pulses and the RF drive is aligned for (a) maximum positive, (b) zero, and (c) maximum negative frequency shift.}
\label{fig:SinefreqshPD}
\end{figure}

Fig.~\ref{fig:SinefreqshPD} show{s} optical and electrical inputs to the spectral-shearing phase modulator for the (a) Maximum positive, (b) zero, and (b) maximum negative frequency shifts for the sine wave at 952~MHz, respectively. {Fig.~\ref{fig:SinefreqshPD} illustrates the relationship between the frequency shift and the slope. In Fig.~\ref{fig:SinefreqshPD}(a), the maximum shift corresponds to the pulse centered on the waveform's maximum positive slope, and similarly there is good correspondence between waveform slope and frequency shift for the other two figure panels with the pulses situated at the zero slope (peak) and maximum negative slope; the corresponding figure for the triangle wave is in the Appendix.} 

\begin{figure}
\centerline{\includegraphics[width=1\columnwidth]{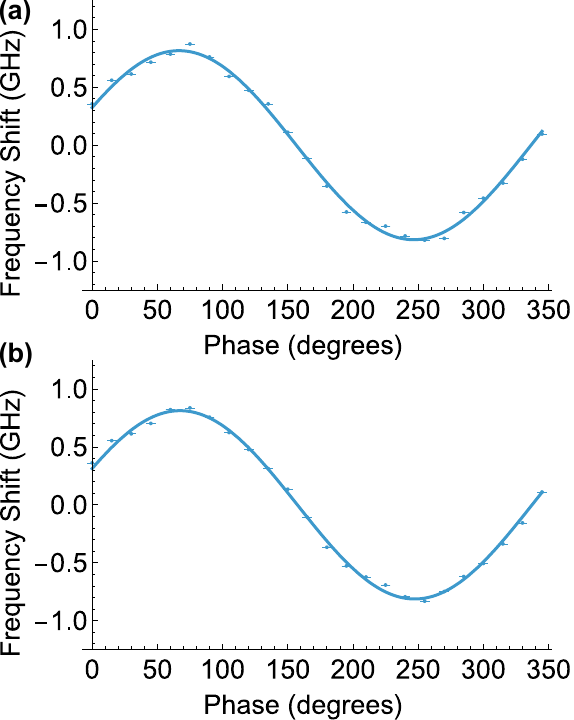}}
\caption{Measured triangle-wave spectral-shearing-induced frequency shift. Frequency shift, on the signal traversing the (a) short and (b) long paths, due to triangle-wave drive vs waveform phase shift. {The 0-GHz shift frequency is calculated from the fitted offset parameter. This parameter fitted to (a) 195.993754~THz and (b) 195.993767~THz.}}
\label{fig:Trigfreqshfit}
\end{figure}
\begin{figure}
\centerline{\includegraphics[width=1\columnwidth]{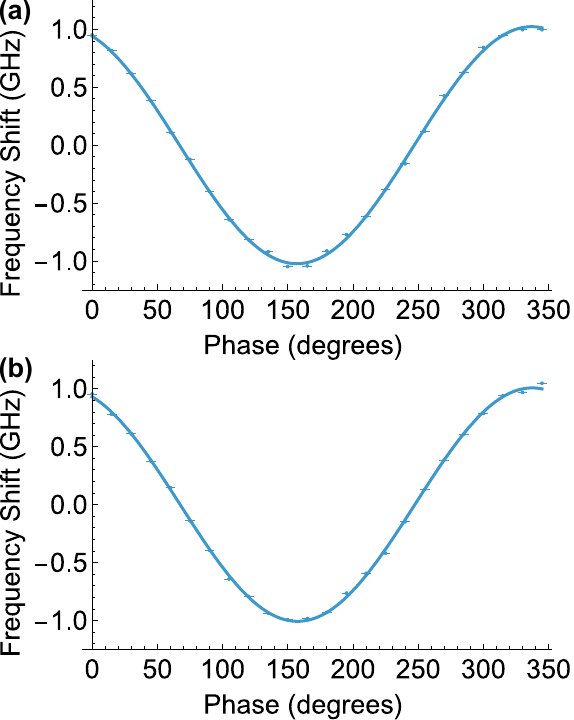}}
\caption{Measured sine-wave spectral-shearing-induced frequency shift. Frequency shift, on the signal traversing the (a) short and (b) long paths, due to sine-wave drive vs waveform phase shift. {The 0-GHz shift frequency is calculated from the fitted offset parameter. This parameter fitted to (a) 195.993754~THz and (b) 195.99376~THz.}}
\label{fig:Sinefreqshfit}
\end{figure}

Due to the bandwidth and sensitivity limitations of the AWG and optical spectrum analyzer (OSA), respectively, we did not examine the sawtooth wave. {To illustrate the effect of using an AWG without sufficient bandwidth to produce a triangle wave at the required frequency, we present Fig.~\ref{fig:Trigfreqshfit} and Fig.~\ref{fig:Sinefreqshfit}.} The derivative of a triangle wave is a square wave, i.e., a true triangle wave just has two different slopes that it switches between. In Fig.~\ref{fig:Trigfreqshfit}, the frequency shift vs triangle-wave phase is plotted and fitted to a sine curve. The fit has $1-R^2<10^{-13}$ which is a very good fit to a sine wave. The missing square wave shape in the spectral shearing and the good fit to a sinewave corroborates the insufficient bandwidth of this waveform generator needed for the triangle wave. Whereas in Fig.~\ref{fig:Sinefreqshfit}, for the sinewave, we find a very good fit to a cosine as expected ($1-R^2=10^{-14}$) and for a larger overall shift compared to the triangle wave. {It is expected that the shift using the triangle wave would be smaller because when the AWG is attempting to emit a triangle wave, the power into the fundamental harmonic is lower leading to a lower overall peak-peak voltage compared to the sinewave at the same frequency.} Fig.~\ref{fig:Trigfreqshfit} and Fig.~\ref{fig:Sinefreqshfit} demonstrate a significant practical advantage of the sinewave over more complex waveforms {and the precise control we have of the generated sign and magnitude of the frequency shift using this technique in our system.}  The sinewave is a pure single frequency tone whereas the triangle wave and sawtooth waves have many Fourier components which can increase the necessary analog bandwidth of the required equipment by $>$10x. {These measurements also illustrate some of the pitfalls possible when attempting to use an instrument to drive frequency shearing without having the sufficient bandwidth for the waveform and frequency chosen while also highlighting some of the advantages of using a simple sinewave.}

In Fig.~\ref{fig:Sinefreqshfit}, the fitted amplitude, corresponding to the maximum shift, is 1.007~GHz and 1.02~GHz for the long and short path, respectively, which is nearly identical as desired for time bins. By using Eq.~\eqref{eq:DelfS} for our experimental parameters, we estimate about 1.2-1.5~GHz shift. The difference is likely due in part to (1) pulse width being appreciable compared to the duration of the shearing waveform, (2) unaccounted for additional insertion loss into the modulator and frequency-dependence in $V_{\pi}$, and (3) some averaging of the center-frequency drift of the laser which was about 100~MHz. 

\begin{figure}
\centerline{\includegraphics[width=1\columnwidth]{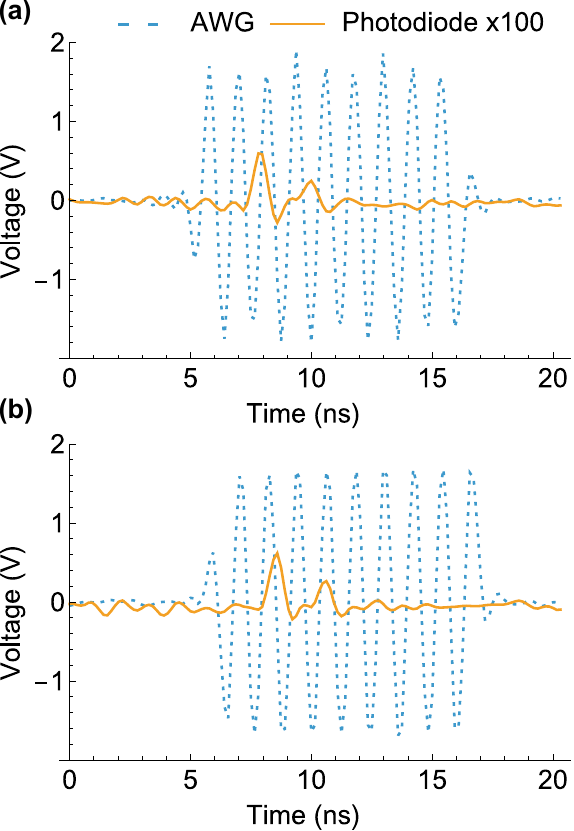}}
\caption{Measured classical time-bin signals entering spectral shearing phase modulator for the 833-MHz sine wave shearing waveform. The phase between the optical pulses and the RF drive is aligned for (a) maximum positive differential, (b) maximum negative differential frequency shift.}
\label{fig:833MSinefreqshPD}
\end{figure}

Having established and characterized our ability to apply a controllable frequency shift via spectral shearing, we turn our attention to characterizing the compatibility of spectral shearing with time-bin pulses. For this, we introduce another waveform in Fig.~\ref{fig:833MSinefreqshPD} which shows a sine wave at 833~MHz whose frequency is chosen such that one time bin is approximately on the maximum slope (e.g., rising edge) and the other time bin is at zero slope (e.g., a peak). This waveform is chosen to illustrate the deleterious phase shift from not applying the same frequency shift to both time bins.

{Our compatibility testing of frequency shearing on time bins included tests} with several waveforms, after applying a frequency shift to the time bins {using a given waveform}, the {time bins} propagate to an analyzer time-bin interferometer. The output power of which is measured and used for time-bin phase stabilization.

\begin{figure}
\centerline{\includegraphics[width=1\columnwidth]{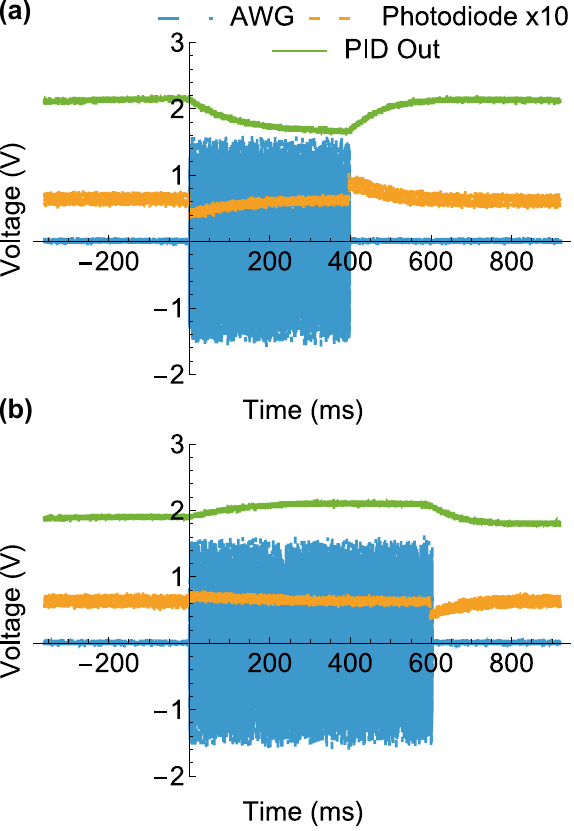}}
\caption{Measured time-bin phase shifts due to 833-MHz sine-wave spectral shearing which applies a different frequency shift to each time bin. The spectral-shearing phase between the optical pulses and the RF drive is aligned for (a) maximum positive differential shift, (b) maximum negative differential frequency shift.}
\label{fig:833onoff}
\end{figure}

\begin{figure}[t!]
\centerline{\includegraphics[width=1\columnwidth]{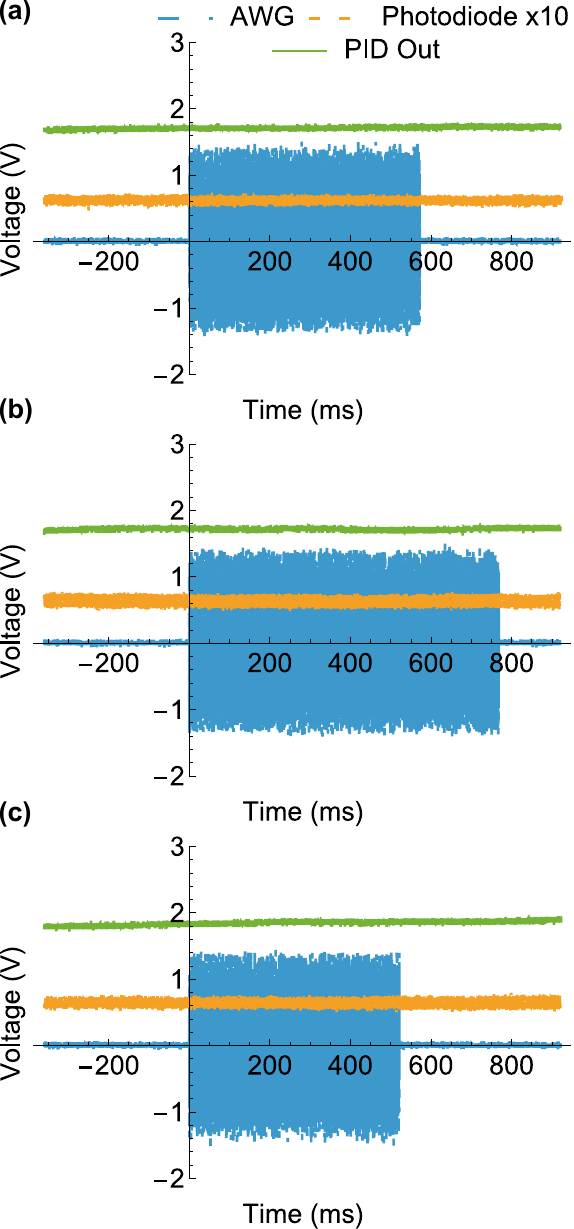}}
\caption{Measured time-bin phase shifts due to 952-MHz sine-wave spectral shearing. The spectral-shearing phase between the optical pulses and the RF drive is aligned for (a) maximum positive, (b) zero, and (c) maximum negative frequency shift.}
\label{fig:952onoff}
\end{figure}

Fig.~\ref{fig:833onoff} shows the result of temporarily applying the 833-MHz spectral shearing waveform to the time bins when the phase stabilization is enabled. For the 833-MHz waveform, we expect there to be a phase shift applied between the time bins since each time bin receives a different frequency shift. Fig.~\ref{fig:833onoff} does indeed show that there is a time-bin phase shift applied by the shearing using the 833-MHz signal evidenced by the changes in the photodiode (PID input) and PID output signals to sense and compensate for the phase shift, respectively. Here we can use the average phase shift between spectral shearing on and off because the spectral shearing in this setup is designed to give a constant shift to every pulse. The PID out trace takes some time to respond to this impulse phase shift due to the integration-term time constant seen in exponential curve in the PID out trace. Overall, this is a good control measurement to show that the undesirable {e}ffect is observable with this measurement in a controlled way. Based on an open-loop calibration of phase shifts measured by the photodiode and induced by PID-out signal, we calculate that the observed average phase shift between the 833-MHz spectral shearing being off and on is about $90^{\circ}$ and ${-}40^{\circ}$ for Fig.~\ref{fig:833onoff}(a) and Fig.~\ref{fig:833onoff}(b), respectively. {A difference is to be expected since we did not precisely calibrate these waveforms to provide exactly equal and opposite phase shifts. Instead, we approximately calibrated them to primarily provide a shift to just one pulse; each waveform was calibrated to provide a shift of a different sign. The purpose of these 833-MHz waveforms is to show in Fig.~\ref{fig:833onoff}(a) and Fig.~\ref{fig:833onoff}(b) that there is a measurable time-bin phase shift due to the mismatched frequency shearing between the time bins. Indeed, this phase shift is obvious and measurable.}

In contrast, changing the drive frequency produces significantly different results. When using a multiple of the inverse time-bin spacing (952~MHz as in Fig.~\ref{fig:SinefreqshPD}), the conditions are set to provide the most identical frequency shift for each time bin as shown in Fig.~\ref{fig:Sinefreqshfit}. Thus, we hypothesize this waveform will lead to no phase shift between the time bins due to the spectral shearing. Fig.~\ref{fig:952onoff} shows the measurements of temporarily applying a 952-MHz sine wave to apply spectral shearing to both time bins. In all cases (maximum, zero, and minimum) frequency shifts, we see no {appreciable} average frequency shift in contrast to  Fig.~\ref{fig:833onoff} where there was clearly a phase shift seen in the PID Out trace. {In the maximum and minimum cases, there is a slight change of the photodiode average value indicating there was a slight phase shift. We ascribe this to a slight imperfection in the calibration of our frequency shearing frequency matching a multiple of the inverse time bin spacing.} From these measurements, we conclude that, under the right circumstances, phase-sensitive time-bin encoding is compatible using spectral shearing to apply a frequency shift to the center frequency of the time-bin pulses.

\section{Discussion}
\label{sec:disc}
Here we propose a heralded frequency-multiplexed zero-added-loss source of time-bin entanglement for which we describe a detailed experimental design. The frequency multiplexing is enabled by spectral shearing. We provide an optimization analysis for spectral shearing used in this application. Moreover, we also show experimentally that spectral shearing is compatible with time-bin qubits which are our chosen qubit type in our proposed heralded entanglement source.

Instead of sending the source outputs for tomographic analysis, use in a quantum repeater protocol would likely involve loading this heralded entanglement into a quantum memory. This has been analyzed in the context of a ZALM source being loaded into a vacancy-based memory~\cite{PhysRevAppliedZALM2023,PhysRevAppliedZALM2024}. This proposal and optimization analysis are generally compatible with narrower frequency bins which would likely be required for a source matched to a quantum memory. 

Our spectral-shearing optimization analysis is conservative in the amount of multiplexing possible, especially in the distinguishability of the frequency bins and the spectral-shearing drive frequency $D$ constraint based on the pulse width. This is to ensure high fidelity discrimination of the frequency bins and uniform spectral shearing applied to each time bin pulse. These constraints are based on the TBP which we have included but a more sophisticated approach to incorporating the TBP into Eq.~\eqref{eq:DelFb} could enable further multiplexing. Additionally, the upper bound of the spectral shearing drive frequency $D$ could be tightened at the expense of reduced uniformity of the spectral shearing slope across the pulse. Currently, the upper bound is set to enforce the phase ramp to be at least eight Gaussian standard deviations wide which provides plenty of time for the slope to be more uniform around the center of the ramp where the pulse is located and ensures the tails also see that uniform slope. To maximize our measured experimental frequency shift, we set $D_S=2/~\Delta t_b>3/2~\Delta t_b$ and did not experience issues in our testing. Moreover, new analysis shows that relaxing the spectral heralding constraints can even further increase the entanglement rate and allow for reduced bin requirements~\cite{shapiroZALM2025}. 

\section{Acknowledgments}
We acknowledge Hsuan-Hao Lu for the inspiration to use spectral shearing as the frequency shift mechanism. This work was performed at Oak Ridge National Laboratory, operated by UT-Battelle for the U.S. Department of energy under contract no. DE-AC05-00OR22725. Funding was provided by the U.S. Department of Energy, Office of Science, Office of Advanced Scientific Computing Research, through
the Performance Integrated Quantum Scalable Internet program (ERKJ432).

J.C.C., M. A., and J. P. designed the proposed ZALM source. J.C.C. developed the optimization analysis. J.C.C. and M.A. constructed time-bin and frequency shearing test setup. J.C.C. devised measurements and analysis for experiment. J.C.C. collected all measurements. J.C.C. analyzed the experimental data. J.P. developed noise analysis. S.G. and N.R. supervised the work. All authors contributed to manuscript preparation.
\appendix

\section{Time-bin and spectral shearing testing} \label{app:tbsetupdesc}
\begin{figure}
\centerline{\includegraphics[width=1\columnwidth]{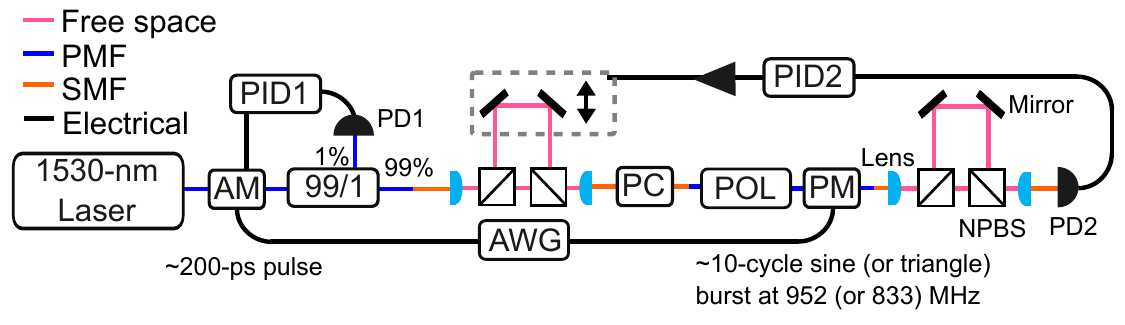}}
\caption{Spectral shearing and time bin compatibility testing experimental setup. Definitions (in alphabetical order). AM: amplitude modulator. AWG: arbitrary waveform generator. NPBS: non-polarizing beamsplitter. PC: polarization controller. PD: photodiode. PID: proportional-integral-derivative controller. PM: phase modulator. PMF: polarization-maintaining fiber. POL: polarizer. SMF: single-mode fiber.}
\label{fig:TBtestsetup}
\end{figure}

{
\subsection{Experimental setup}
}
To experimentally test the application of spectral shearing to time-bins, we constructed the setup shown in Fig.~\ref{fig:TBtestsetup}. The output of a CW 1529.6-nm laser (Pure Photonics PPCL590) is directed towards an amplitude modulator (EOSpace AX-0MSS-10-PFA-PFAP) for pulse carving by an arbitrary waveform generator (Berkeley Nucleonics Corp. Model 685-2C-SE2) which emits an approximately 200-ps FWHM pulse. The modulator DC bias is stabilized using a 99/1 polarization-maintaining pick-off (AC Photonics) detected by a power meter (Thorlabs PM101A) used as an error signal by a linear controller (Liquid Instruments Moku:GO PID instrument) to adjust the modulator bias to maintain stable output power. 

The resulting laser is then sent to a free-space unbalanced Mach-Zehnder interferometer (UMZI) which creates the time-bins. The output is coupled back into single-mode fiber and transferred to polarization-maintaining fiber via polarization controller and polarizer where it is then directed to a phase modulator (EOSpace PM-DVEK-40-PFA-PFA-LV-UL). The phase modulator is driven by the other channel of the aforementioned arbitrary waveform generator. The waveform is nominally 5~V$_\text{pk-pk}$ and is changed between sine and triangle shape. We program a burst of 10 cycles of 952.381~MHz (based on the UMZI delay). This frequency slightly exceeds our upper-bound for the spectral shearing in Eq.~\eqref{eq:DTub} and Eq.~\eqref{eq:DSub} but that is a pretty conservative upper-bound to make sure the tails of the pulse are included so we do not notice issues going above it by $2/1.5=1.333$. We also used a waveform with several cycles of 833~MHz $=\frac{1.75}{2}(952\text{ MHz})$ whose frequency is chosen such that one time bin is approximately on the maximum slope (e.g., rising edge) and the other time bin is at zero slope (e.g., a peak).

The output of the phase modulator is sent to another UMZI for time-bin analysis and is detected by a power meter (Thorlabs PM100D with head S122C). In this detected signal, we do see some additional noise due to interference between the paths of the first time-bin interferometer. The analog output is directed to the other channel of the aforementioned linear controller to implement piezo-based phase stabilization by moving a translation stage on the first long arm using a piezo-electric crystal and driver (Thorlabs KPZ101). This configuration enabled testing for phase shifts created between the time-bins due to the spectral shearing. To trigger the oscilloscope (Agilent MSO-X 4104A) for data acquisition of the measurements shown in Fig.~\ref{fig:833onoff} and Fig.~\ref{fig:952onoff}, we leveraged the external termination of the phase modulator to connect the output of the phase modulator to a 50-$\Omega$-terminated oscilloscope channel.

For some measurements, we also redirected the optical and electrical signals going to the phase modulator to instead goto an oscilloscope (Agilent MSO-X 4104A) with the optical signal first being detected by a GHz-bandwidth photodiode (Thorlabs DET08CFC). This configuration, enabled comparing the optical and electrical inputs to the spectral shearing stage.

\begin{figure}
\centerline{\includegraphics[width=1\columnwidth]{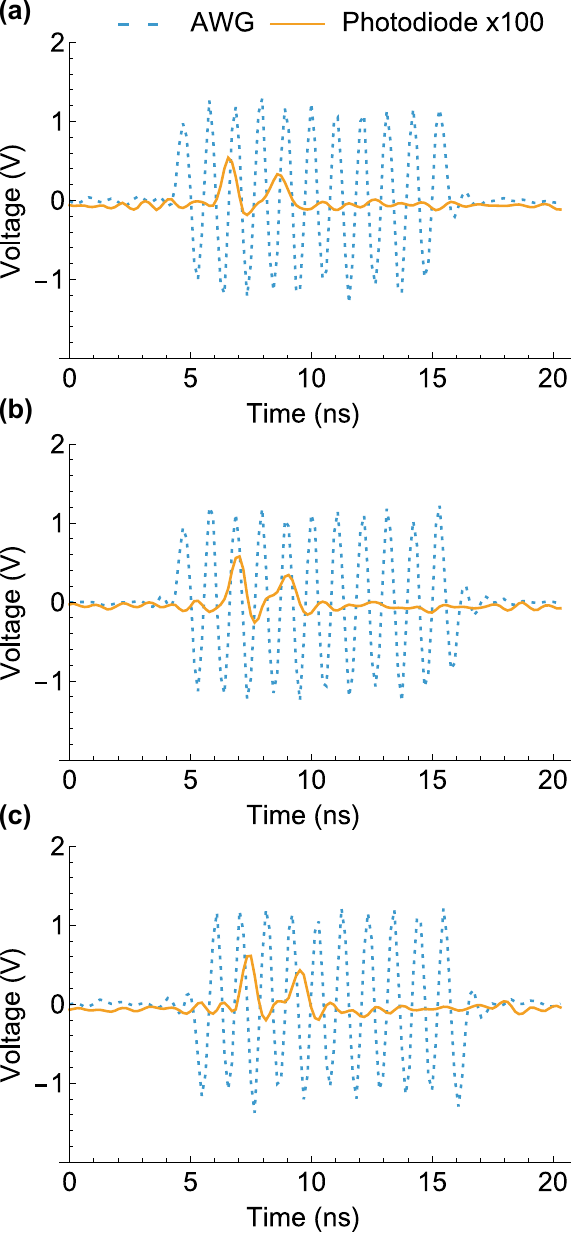}}
\caption{Measured classical time-bin signals entering spectral shearing phase modulator for the triangle shearing waveform. The phase between the optical pulses and the RF drive is aligned for (a) maximum positive, (b) zero, and (c) maximum negative frequency shift.}
\label{fig:TrigfreqshPD}
\end{figure}

To measure the spectral shearing amount, the light exiting the phase modulator was sent to a spectrum analyzer. We extracted the mean of a Gaussian fit of the measured spectrum from a 7.5-GHz resolution OSA (Thorlabs OSA202C).
{
\subsection{Additional test results}
}
{Fig.~\ref{fig:TrigfreqshPD} shows optical and electrical inputs to the spectral-shearing phase modulator for the (a) Maximum positive, (b) zero, and (c) maximum negative frequency shifts for the triangle wave at 952~MHz, respectively.} 
The shearing waveforms  {in Fig.~\ref{fig:TrigfreqshPD} and Fig.~\ref{fig:SinefreqshPD}} look fairly similar indicating that the triangle wave needs more analog bandwidth to completely have the desired shape on the tips of the waveform. 
\clearpage
%

\end{document}